\documentclass[twocolumn,aps,superscriptaddress,showpacs,nofootinbib,floatfix]{revtex4-1}
\usepackage{listings}
\usepackage[utf8]{inputenc}
\usepackage{hyperref}
\usepackage{amsmath}
\usepackage{amssymb}
\usepackage{amsfonts}
\usepackage{tabularx}
\usepackage{booktabs}
\usepackage{graphicx}
\usepackage{color}
\usepackage{xcolor}
\usepackage{float}
\usepackage{multirow}
\usepackage[inline]{enumitem}
\usepackage[T2A,T1]{fontenc}

\graphicspath{{fig/}}
\definecolor{theblue}{RGB}{0,50,230}

\usepackage{hyperref}
\hypersetup{
  colorlinks=true,
  linkcolor=theblue,
  citecolor=theblue,
  urlcolor=theblue
}

\newcommand {\avg}[1]{\ensuremath{\langle\kern-1.0pt\langle#1\rangle\kern-1.0pt\rangle}}

\newlength\cmsFigWidth

\usepackage[normalem]{ulem}  
\usepackage{soul, xcolor}

\renewcommand\sout{\bgroup \color{red} \ULdepth=-.5ex \ULset}

\begin{document}

\title{From hydro to jet quenching, coalescence and hadron cascade: a coupled approach to solving the $R_{AA}\otimes v_2$ puzzle}

\author{Wenbin~Zhao}
\affiliation{Key Laboratory of Quark and Lepton Physics (MOE) \& Institute of Particle Physics,Central China Normal University, Wuhan 430079, China}
\author{Weiyao Ke}
\affiliation{Physics Department, University of California, Berkeley, California 94720, USA}
\affiliation{Nuclear Science Division MS 70R0319, Lawrence Berkeley National Laboratory, Berkeley, California 94720, USA}
\author{Wei Chen}
\affiliation{School of Nuclear Science and Technology, University of Chinese
Academy of Sciences, Beijing 100049, China}
\author{Tan Luo}
\affiliation{Instituto Galego de F\'isica de Altas Enerx\'ias IGFAE, Universidade de Santiago de Compostela, E-15782 Galicia-Spain}
\author{Xin-Nian Wang}
\email[]{xnwang@lbl.gov, current address $^3$.}
\affiliation{Key Laboratory of Quark and Lepton Physics (MOE) \& Institute of Particle Physics,Central China Normal University, Wuhan 430079, China}
\affiliation{Nuclear Science Division MS 70R0319, Lawrence Berkeley National Laboratory, Berkeley, California 94720, USA}

\date{\today}

\begin{abstract}

Hydrodynamics and jet quenching are responsible for the elliptic flow $v_2$ and suppression of large transverse momentum ($p_T$) hadrons, respectively, two of the most important phenomena leading to the discovery of a strongly coupled quark-gluon plasma (QGP) in high-energy heavy-ion collisions.  A consistent description of the hadron suppression factor $R_{AA}$ and $v_2$, especially at intermediate $p_T$, however, remains a challenge.  We solve this long-standing $R_{AA}\otimes v_2$ puzzle by including quark coalescence for hadronization and final state hadron cascade in the coupled linear Boltzmann transport-hydro model that combines concurrent jet transport and hydrodynamic evolution of the bulk medium. We illustrate that quark coalescence and hadron cascade, two keys to solving the puzzle, also lead to a splitting of $v_2$ for pions, kaons and protons in the intermediate $p_T$ region. We demonstrate for the first time that experimental data on $R_{AA}$, $v_2$ and their hadron flavor dependence from low to intermediate and high $p_T$ in high-energy heavy-ion collisions can be understood within this coupled framework.
\end{abstract}

\maketitle 

\noindent{\it 1. Introduction}: 
Experimental evidences at the Relativistic Heavy-Ion Collider (RHIC) and the Large Hadron Collider (LHC) have confirmed the existence of a strongly coupled quark-gluon plasma (QGP) in high-energy heavy-ion (A+A) collisions \cite{Gyulassy:2004zy,Adams:2005dq,Adcox:2004mh,Jacobs:2004qv,Muller:2012zq}. These include strong anisotropic flow at low transverse momentum ($p_T$) \cite{Ackermann:2000tr,Adcox:2002ms,Aamodt:2010pa,ATLAS:2011ah,Chatrchyan:2012wg} and suppression of hadrons at high $p_T$ \cite{Adcox:2001jp,Adler:2002xw,Wang:2004dn,Aamodt:2010jd,Majumder:2010qh}. The spectra and flow patterns of bulk hadrons at low $p_T\lesssim 2$ GeV/$c$ are well described by the hydrodynamic expansion of the QGP as a strongly-coupled fluid~\cite{Romatschke:2007mq,Song:2010mg,Gale:2012rq,Huovinen:2013wma,Heinz:2013th,Gale:2013da,Niemi:2015qia,Bernhard:2016tnd,McDonald:2016vlt,Zhao:2017yhj}. At high $p_T \gtrsim 10$ GeV/$c$, the hadron suppression factor ($R_{AA}$) and the azimuthal anisotropy ($v_2$) \cite{Adams:2003am,Sirunyan:2017pan,Aaboud:2018ves,Acharya:2018lmh,Xing:2019xae,Liu:2021izt} can be quantitatively understood in terms of jet quenching caused by parton energy loss as hard partons propagate through the QGP medium \cite{Wang:1991xy,Wang:1998ww,Vitev:2002pf,Wang:2002ri,Eskola:2004cr,Qin:2007rn,Schenke:2009gb,Chen:2011vt,Majumder:2011uk,Buzzatti:2011vt,Zapp:2012ak,Gyulassy:2000gk,Kumar:2017des,Zigic:2019sth, Mehtar-Tani:2013pia,Qin:2015srf,Blaizot:2015lma}. 
In the intermediate $p_T \sim 2 - 10$ GeV/$c$ region where soft and hard physics interface it remains, however, a challenge to describe the hadron spectra and the azimuthal anisotropy consistently. There is a longstanding puzzle that parton energy loss models that are adjusted to describe hadron $R_{AA}$ under-predict the azimuthal anisotropy $v_2$~\cite{Molnar:2013eqa, Noronha-Hostler:2016eow, Zhang:2013oca, Liao:2008dk, Kopeliovich:2012sc,Cao:2017umt,Andres:2019eus} at intermediate $p_T$. Parton energy loss alone also cannot describe the constituent quark number (NCQ) dependence of hadron $R_{AA}$ and $v_2$.

Many attempts tried to solve this puzzle, from evoking the exotic mono-poles in interactions between the hard partons and the medium 
near the pseudo-critical temperature $T_c$ \cite{Xu:2014tda,Shi:2018lsf,Shi:2019nyp} to taking into account of the event-by-event fluctuation of the bulk medium \cite{Noronha-Hostler:2016eow} and combining hydrodynamics with jet transport as in EPOS \cite{Werner:2012xh} and HYDJET++ model \cite{Lokhtin:2012re}. While such an exotic interaction with a drastically large jet-medium coupling at $T_c$ is not needed to describe $R_{AA}$ and $v_2$ of full jets \cite{Aad:2013sla,Zapp:2013zya,Adam:2015mda,He:2019zld} in which hadron's flavor information is not considered, the event-by-event fluctuation of the bulk medium is found not to significantly increase 
high $p_T$ $v_2$ \cite{Betz:2011tu,Cao:2017umt,He:2019zld}. Furthermore, none of these attempts can address the NCQ dependence of $R_{AA}$ and $v_2$ in the intermediate $p_T$ region.

It is well known that quark coalescence is the key to explain the observed NCQ scaling of hadronic anisotropies and the enhanced baryon to meson ratios in A+A relative to proton-proton ($p+p$) collisions at intermediate $p_T$  \cite{Fries:2003vb,Fries:2003kq,Greco:2003xt,Greco:2003mm,Greco:2003vf,Molnar:2003ff,Hwa:2004ng,Adams:2004bi,Adams:2005zg,Adare:2006ti,Abelev:2007ra,Adare:2012vq,Adare:2011vy}. Furthermore, it has been shown that hadrons, especially baryons, at intermediate $p_T$ are also sensitive to rescatterings in the hadronic phase \cite{Shen:2010uy,Ryu:2017qzn,Knospe:2021jgt}. Quark coalescence in a hydrodynamic medium combined with parton transport, fragmentation and hadron cascade should, therefore, have the potential to solve the $R_{AA}\otimes v_2$ puzzle and describe hadron production from low to intermediate and high $p_T$ in high-energy heavy-ion collisions. 

In this work, we implement the quark coalescence, especially between thermal and jet shower partons, and a hadronic afterburner in the state-of-the-art coupled linear Boltzmann transport(CoLBT)-hydro model \cite{Chen:2020tbl} that has concurrent evolution of both the bulk medium and jet showers, including jet-induced medium responses. We carry out a first study that couples event-by-event hydrodynamics, jet quenching, quark coalescence and hadron cascade. We demonstrate that this fully coupled approach can simultaneously describe $R_{AA}$, differential $v_2$, and their NCQ dependence in the full range of $p_T$ in high-energy heavy-ion collisions, therefore solving the longstanding $R_{AA}\otimes v_2$ puzzle that connects the two most important aspects of the discovery of a strongly coupled QGP. This also sheds light on the hadronization mechanism of QGP in high-energy heavy-ion collisions.

\noindent{\it 2.} CoLBT-{\it hydro model}: The CoLBT-hydro model~\cite{Chen:2017zte,Chen:2020tbl,Yang:2021qtl} is developed to simulate the concurrent evolution of jet showers and the bulk medium by coupling the (3+1)D 
Central China Normal University and the Lawrence Berkeley National Laboratory (CCNU- LBNL) viscous hydrodynamic model and with OpenCL GPU parallelization (CLVISC)~\cite{Pang:2012he,Pang:2018zzo} with the linear Boltzmann transport (LBT) model~\cite{Wang:2013cia,Li:2010ts, He:2015pra,Cao:2016gvr,Cao:2017hhk}. The LBT model treats the propagation of jet shower and thermal recoil partons on an equal footing and includes both pQCD elastic scattering and medium-induced gluon radiation within the high-twist approach \cite{Guo:2000nz,Wang:2001ifa,Zhang:2003yn,Schafer:2007xh}. The coupling between LBT and CLVisc is through an energy-momentum source term deposited by soft partons in the hydrodynamic equation,
\begin{eqnarray}
& & \partial_{\mu}T_{\rm fluid}^{\mu\nu}=J^{\nu},
\label{eq:hydro} \\
& & J^{\nu}=\sum_{i} \frac{\theta(p^{0}_{\rm cut}-p_{i}\cdot u)p^{\nu}}{\tau(2\pi)^{3/2}\sigma^{2}_{r}\sigma_{\eta_{s}}\Delta\tau}
e^{-\frac{(\vec{x}_{\perp}-\vec{x}_{\perp i})^{2}}{2\sigma^{2}_{r}}
-\frac{(\eta_{s}-\eta_{si})^{2}}{2\sigma^{2}_{\eta_{s}}}},
\label{eq:smear}
\end{eqnarray}
where $T_{\rm fluid}^{\mu\nu}$ is the energy-momentum tensor of the bulk medium, the summation in the Gaussian smearing in the Milne coordinates is over both soft recoil, radiated  and ``negative'' partons --``hole particles'' created in elastic jet-medium collisions. 
For studies presented in this Letter, we set $p^0_{\rm cut}=3.0 $ GeV/$c$, $\sigma_r$=0.6 fm and $\sigma_{\eta_s}$=0.6. 
Jet-induced medium response is found essential to describe many jet observables in heavy-ion collisions, such as jet shape \cite{Luo:2018pto}, jet fragmentation function \cite{Chen:2020tbl}, $\gamma/Z$-hadron correlations \cite{Chen:2017zte,Yang:2021qtl} and baryon-to-meson ratio in and around jet \cite{Chen:2021rrp,Luo:2021voy}. However, it has negligible influence on low $p_T$ single inclusive hadron spectra due to the dominance of soft particles from the bulk medium. 

In the CLVisc hydrodynamic evolution, a lattice QCD inspired equation of state~\cite{Bazavov:2014pvz} is used. The TRENTo model~\cite{Moreland:2014oya} with optimized parameters ~\cite{[][{. Initial entropy normalization factor is 140,  the reduced-thickness parameter $p=0$, entropy despoliation fluctuation parameter is $k=1.5$, and the nucleon width parameter is $w=0.8$ fm}.]Bernhard:2016tnd}
and a longitudinal envelope function~\cite{[][{The plateau width is $\eta_{\rm flat}=2.0$ fm and the Gaussian fall-off width is $\sigma_{\eta}=1.8$ fm}.]envelop,Pang:2012he,Pang:2018zzo} is used to generate the initial entropy density profile with event-by-event transverse fluctuation. The specific shear viscosity $\eta/s$=0.10, the freeze-out temperature $T_{\rm sw}=150$ MeV, the initial time $\tau_0=0.6$ fm/$c$ and parameters in the initial entropy profile have been adjusted to reproduce the charged hadron multiplicity, $p_T$ spectra and integrated flow harmonics $v_n$ at mid-rapidity in A+A collisions. While a finite starting time $\tau_0$ of the jet-medium interaction is a default assumption in LBT and  CoLBT-hydro, other studies \cite{Andres:2019eus,Zigic:2019sth} find it necessary to achieve $v_2$ at high $p_T$ that is compatible with experimental data. We also assume that hard shower partons free-stream during the formation time $\tau_f = 2z(1-z)E/k_{\perp}^2$ before they interact with the QGP medium, where $k_{\perp}$ is the transverse momentum, $z$ the energy fraction of the shower parton after the initial splitting from its mother parton with energy $E$. Further details about the LBT and CLVisc model can be found in Refs.~\cite{Li:2010ts,Wang:2013cia,He:2015pra,Cao:2016gvr,Cao:2017hhk} and Refs.~\cite{Pang:2018zzo,Pang:2012he}, respectively.

We use PYTHIA8 \cite{Sjostrand:2007gs} with EPPS16 nuclear parton distributions~\cite{Eskola:2016oht} to generate initial jet showers.
A minimum hard scale $\hat p_{T0}=$ 4.0 GeV/$c$ is set for jet production. The average number of such jet production per event is $\left<N_{\rm jet}\right>= \langle N_{\rm coll} \rangle P_{\rm trigger}^{\rm jet}$, where $\langle N_{\rm coll} \rangle$ is the number of binary nucleon-nucleon ($N+N$) collisions whose transverse distribution is given by the nuclear overlap function. $P_{\rm trigger}^{\rm jet}$ is the probability for a minimum-bias $N+N$ collision to have at least one pair of jet production,
\begin{eqnarray}
P_{\rm trigger}^{\rm jet} = \frac{1}{\sigma_{NN}^{\rm inel}}\int d^2b \left[1-e^{-T_{NN}(b)\sigma_{\rm jet}(\hat p_T>\hat p_{T0})}\right],
\end{eqnarray}
where $T_{NN}(b)$ is the nucleon overlapping function and $\sigma_{NN}^{\rm inel}=70$ mb at $\sqrt{s_{NN}} = 5.02$ TeV.


There are two adjustable parameters in this study of single inclusive hadron spectra using the CoLBT-hydro model. While the strong coupling $\alpha_{\rm s}$ for jet production and showering is allowed to run according to pQCD \cite{Cao:2016gvr,Cao:2017hhk}, the effective coupling to the medium partons $\alpha_{\rm s}=0.17$ is adjusted to fit  $R_{AA}$ at high $p_T$ in central Pb+Pb collisions at $\sqrt{s_{NN}} = 5.02$ TeV. The second parameter is a lower cut-off $p_{T{\rm min}}$ for initial jet partons that propagate according to LBT. Initial partons below this scale are assumed to be thermalized as part of the initial condition for the hydro evolution of the bulk medium. The value  $p_{T{\rm min}}=7.0$ (5.0) GeV/$c$ is tuned by fitting the final hadron spectra at intermediate $p_T$ in 10-20\% (40-50\%) central Pb+Pb collisions at $\sqrt{s_{NN}} = 5.02$ TeV. Its value depends on the centrality and colliding energy.

\noindent {\it 3. Hadronization and hadron cascade}: In this study, we adapt the Hydro-Coal-Frag hybrid model~\cite{Zhao:2019ehg} for hadronization that includes hydro freeze-out at low $p_T$, quark coalescence at intermediate $p_T$ and fragmentation at high $p_T$. The interplay between hadron freeze-out in hydro and parton dynamics is defined by a separation scale $p_{Ts}=1.5$ GeV/$c$ for the effective constituent quarks, above which viscous corrections to the equilibrium distribution become large and parton coalescence and fragmentation become the relevant mechanisms for hadronization.  This scale corresponds to
 $p_{T\rm meson} < 3$ GeV/$c$ and $p_{T\rm baryon} < 4.5$ GeV/$c$ for hadron production through hydro freeze-out on the switching hyper-surface. Accordingly,
 thermal quarks with $p_T>$1.5 GeV/$c$ on the switching hyper-surface \cite{[][{In principle, the energy and momentum of the coalesced thermal partons with $p_T<$1.5 GeV/$c$ should be subtracted from the bulk medium to conserve the total energy and momentum. However, we find the effect is negligible in the final hadron spectra, which we will neglect}.]thermal} are allowed to participate in coalescence processes for hadronization, which include thermal-thermal, thermal-shower and shower-shower coalescence. Shower partons include both jet shower and hard medium recoil partons passing through the isothermal hyper-surface in CoLBT-hydro.

Shower partons that do not coalesce will hadronize through string fragmentation using PYTHIA8~\cite{Sjostrand:2007gs} with tuned strangeness suppression~\cite{[{We tune the Lund string parameter {\tt probStoUD}=0.4, which controls the suppression of $s$ quark pair production relative to $u,d$ quarks in the string fragmentation, to reproduce the $K/\pi$ ratio in minimum-bias p-p collisions at $\sqrt{s}=$5.02 TeV~\cite{Acharya:2018orn}.}]strange}. We adopt a colorless hadronization scheme \cite{Kumar:2019bvr} for these shower partons that should have lost their original color configurations and form strings with the distances $\Delta R=\sqrt{(\Delta \eta)^2 + (\Delta \phi)^2}$ of neighboring parton pairs  minimized.
Finally, the Ultra-relativistic Quantum Molecular Dynamics (UrQMD)~\cite{Bass:1998ca, Bleicher:1999xi} model is used to perform hadronic rescatterings and resonance decays in the hadronic stage of the system until kinetic freeze-out.

\begin{figure}[t]
  \includegraphics[scale=0.43]{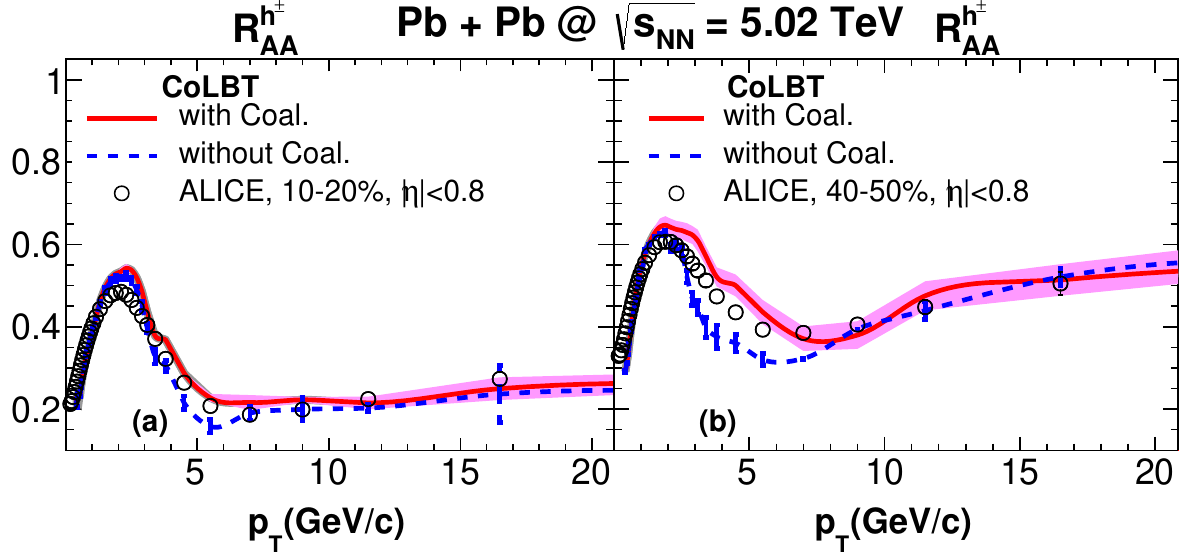}
  \caption{(Color online) The nuclear modification factor $R_{\rm AA}$ of charged hadrons in (a) 10-20\% and (b) 40-50\% Pb+Pb collisions at $\sqrt{s_{NN}}=$5.02 TeV from CoLBT-hydro simulations with (solid) and without quark coalescence(dashed) as compared to ALICE experimental data~\cite{Acharya:2018qsh}.}
 \label{fig:raa}
\end{figure}

\begin{figure}[h]
  \includegraphics[scale=0.43]{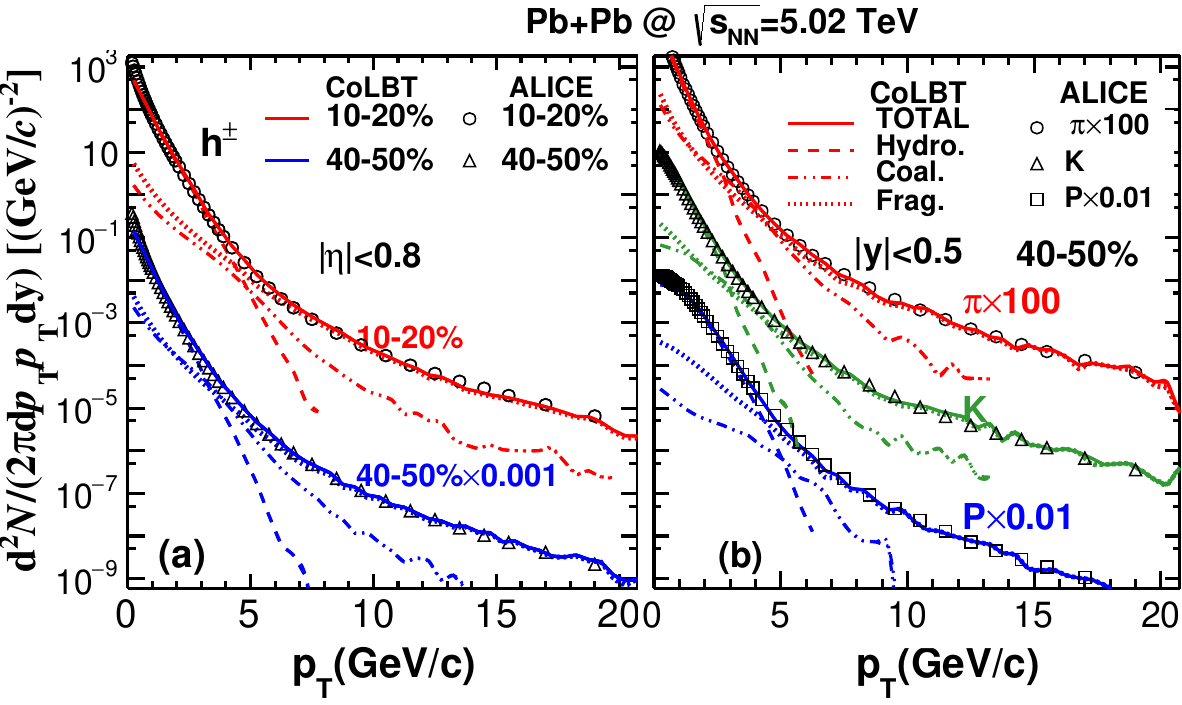}
  \caption{(Color online) Spectra of (a) charged hadrons (solid) in 10-20\% and 40-50\%  and (b) identified $\pi$, $K$ and $p$ (solid) in 40-50\% Pb+Pb collisions at $\sqrt{s_{NN}}=$5.02 TeV and contributions from hydro freeze-out (dashed), parton coalescence (dot-dashed) and fragmentation (dotted) in CoLBT-hydro simulations as compared to ALICE experimental data~\cite{Acharya:2018qsh,Acharya:2019yoi}. }
  \label{fig:spectraall}
\end{figure}

 \noindent {\it 4. Nuclear modification of hadron spectra}: Using CoLBT-hydro with the inclusion of quark coalescence and hadronic afterburner, we calculate hadron spectra from low to intermediate and high $p_T$ in A+A collisions. Shown in Fig.~\ref{fig:raa} are CoLBT-hydro results on the nuclear modification factor $R_{AA}(p_T)$ of charged hadron spectra~\cite{[][{Experimental data on the hadron spectra in p+p collision at $\sqrt{s_{NN}}$=5.02 TeV~\cite{Acharya:2018qsh} are used as the baseline for the model calculations of $R_{AA}$.}]ppdata} which describe reasonably well the experimental data \cite{Acharya:2018qsh} for 10-20\% and 40-50\% Pb+Pb collisions at $\sqrt{s_{NN}}=5.02$ GeV. Only statistical uncertainties in the CoLBT-hydro results are shown as solid bands. To illustrate the different hadron production mechanisms underlying $R_{AA}(p_T)$ in different $p_T$ regions, we show in Fig.~\ref{fig:spectraall} (a)  $p_T$ spectra of charged hadrons in 10-20\% and 40-50\% and (b) identified pion ($\pi$), kaon ($K$) and proton ($p$) spectra from CoLBT-hydro in 40-50\% Pb+Pb collisions as compared to the experimental data \cite{Acharya:2018qsh,Acharya:2019yoi}. Also shown are contributions from hydro (dashed), parton coalescence (dot-dashed) and fragmentation (dotted) in CoLBT results.
 
\begin{figure}[h]
  \includegraphics[scale=0.43]{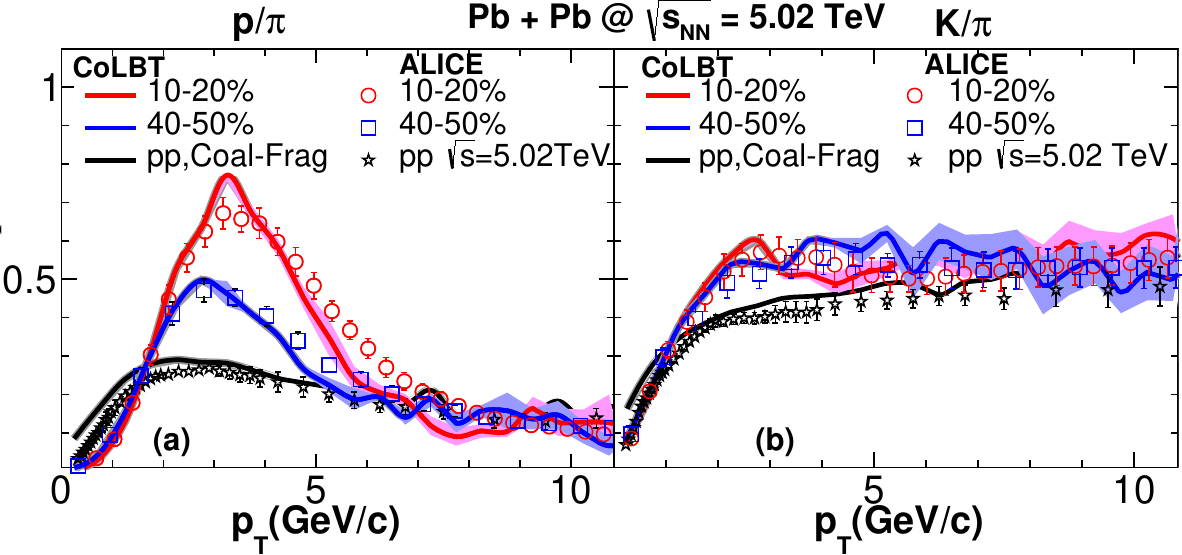}
  \caption   {(Color online) (a) $p/\pi$ and (b) $K/\pi$ ratio in 10-20\%  and 40-50\%  Pb+Pb and $p+p$ collisions at $\sqrt{s_{NN}}=$5.02 TeV from CoLBT-hydro as compared to experimental data~\cite{Acharya:2019yoi}.  }
  \label{fig:ratio}
\end{figure}

 In CoLBT-hydro, hadron spectra are dominated by the hydro contribution below $p_T<2$ GeV/$c$ where $R_{AA}$ increases rapidly with $p_T$ due to the strong radial flow from the hydrodynamic expansion until it peaks at $p_T\approx 2$ GeV/$c$. At intermediate $2<p_T<6$ GeV/$c$, $R_{AA}$ decreases from its peak value due to the onset of contributions from parton coalescence and fragmentation in which hadron spectrum are suppressed due to parton energy loss. At large $p_T>8$ GeV/$c$ where fragmentation prevails,  $R_{AA}$ is determined by the energy dependence of the parton energy loss and the initial jet spectra. 

 We note that the transition in the underlying hadron production mechanism, from hydrodynamics to parton coalescence and fragmentation, occurs at higher $p_T$ in more central collisions because of stronger radial flow, pushing the realm of hydrodynamics to higher $p_T$. The transition also happens at larger $p_T$ for baryons than mesons. This is because of the constituent-quark-based separate scale $p_{Ts}=1.5$ GeV/$c$ that leads to a larger cut-off $p_{T{\rm baryon}}<4.5$ GeV/$c$ for baryons than $p_{T{\rm meson}}<3$ GeV/$c$ for mesons from hydro. The hydro spectra are also mass-ordered due to radial flow. Hydro contributions to the hadron spectra above the $p_T$ cut-off come from hadron cascade within UrQMD during the hadronic evolution. The interplay between hydro, parton energy loss, coalescence and fragmentation implemented in CoLBT-hydro can describe the flavor dependence of hadron spectra and their medium modification as seen in Fig.~\ref{fig:spectraall} (b).

To examine the flavor composition of the hadron spectra and their $p_T$ dependence in detail, we compare the CoLBT-hydro results on $p/\pi$ and $K/\pi$ ratio as a function of $p_T$ to the experimental data in Fig.~\ref{fig:ratio}. Both $p/\pi$ and $K/\pi$ ratio exhibit a steep increase from $p_T=0 - 3$ GeV/$c$ with a mass ordering induced by the radial flow, an intrinsic feature of hydrodynamic models. For $p_T>3$ GeV/$c$, $p/\pi$ decreases with $p_T$ while $K/\pi$ remains constant as a result of the interplay between hydrodynamic expansion, quark coalescence and fragmentation. At large $p_T>8$ GeV/$c$, these hadron ratios in Pb+Pb approach the values in $p+p$ collisions when the hadronization is dominated by vacuum-like parton fragmentation. CoLBT-hydro describes these features in the data well.

\noindent {\it 5. Coalescence and $R_{AA}\otimes v_2$ puzzle}:
To demonstrate the importance of parton coalescence in resolving the $R_{AA}\otimes v_2$ puzzle, we examine first the azimuthal anisotropy $v_2$ of the hadron spectra for different hadron species. Following the experiments~\cite{Sirunyan:2017pan,Acharya:2018lmh,Acharya:2018zuq}, we use the scalar-product (SP) method to compute $v_2$ of charged hadrons within $|\eta|<1.0$, using reference particles in  $|\eta|<1.0$ and 0.2$<p_T<$ 5.0 GeV/$c$. 

\begin{figure}
  \includegraphics[scale=0.43]{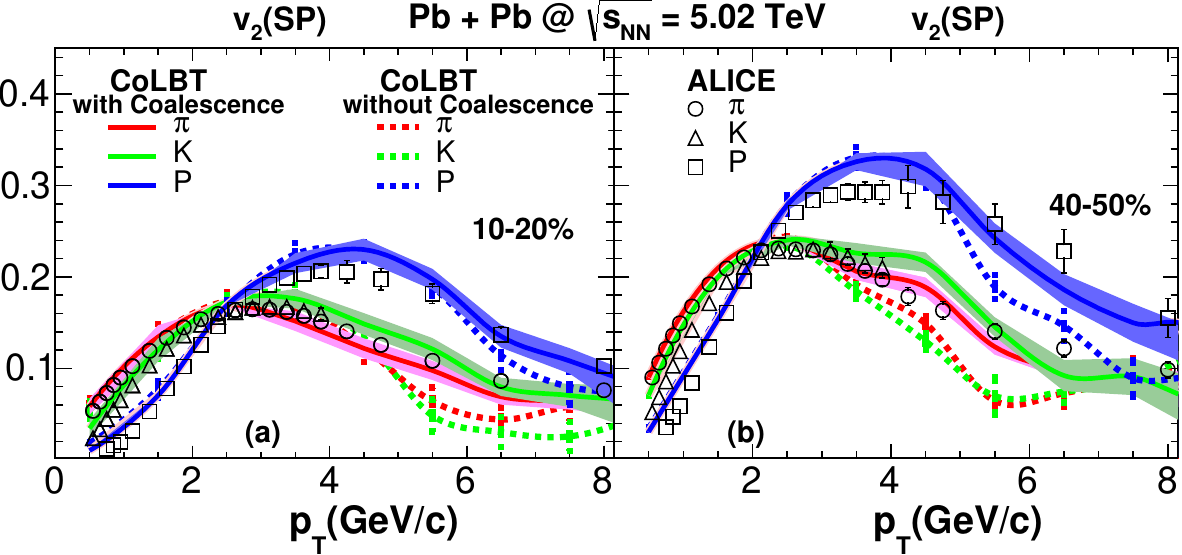}
  \caption{(Color online) CoLBT-hydro results on $v_{2}(p_T)$ for $\pi$, $K$ and $p$  in (a) 10-20\% and (b) 40-50\% Pb+Pb collisions at $\sqrt{s_{NN}}=$5.02 TeV with (solid) and  without (dashed) contributions from quark coalescence compared to ALICE data~\cite{Acharya:2018zuq}.  }
  \label{fig:v2pid1020}
\end{figure}
\begin{figure}
  \centering \includegraphics[scale=0.31]{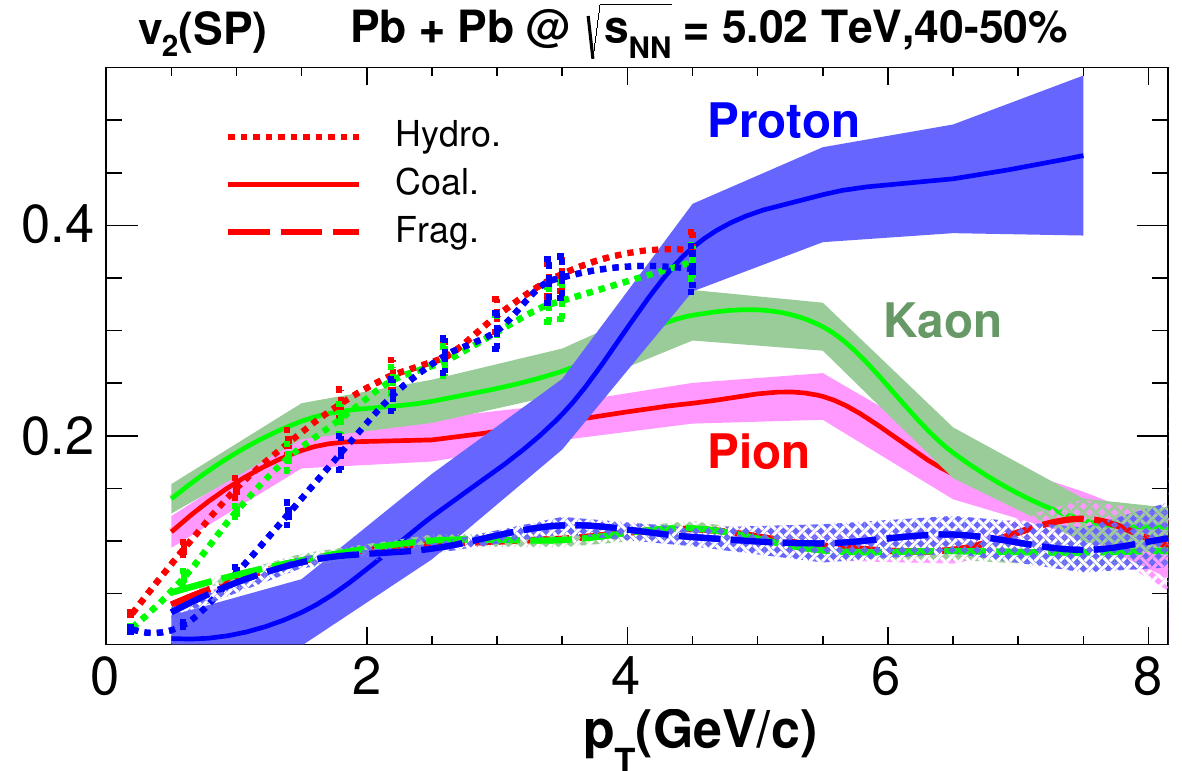}
  \caption{(Color online) CoLBT-hydro results on $v_{2}(p_T)$ for $\pi$, $K$ and $p$ from hydro freeze-out (dotted), parton coalescence (solid) and fragmentation (dashed) in 40-50\%  Pb+Pb collisions at $\sqrt{s_{NN}}=$5.02 TeV.}
  \label{fig:v2ptcontribution}
\end{figure}

Shown in Fig.~\ref{fig:v2pid1020} are $v_2(p_T)$ for $\pi$, $K$ and $p$ from CoLBT-hydro with (solid) and without (dashed) contributions from coalescence in 10-20\% and 40-50\% Pb+Pb collisions as compared to the experimental data \cite{Acharya:2018zuq}. The interplay among hydro, 
jet quenching, coalescence and fragmentation in $v_2(p_T)$ is very similar to that in $R_{AA}(p_T)$. To help understand this interplay, we also plot in Fig.~\ref{fig:v2ptcontribution} $v_2(p_T)$ for $\pi$, $K$ and $p$  from hydro freeze-out (dotted), parton fragmentation (dashed) and quark coalescence (solid) which has a strong flavor dependence.

The rapid increase and the mass ordering of $v_2(p_T)$ from hydro at low $p_T<2$ GeV/$c$ in Figs.~\ref{fig:v2pid1020} and \ref{fig:v2ptcontribution} are characteristics of viscous hydrodynamics with hadronic afterburner~\cite{Huovinen:2013wma,Heinz:2013th,Gale:2013da,McDonald:2016vlt, Zhao:2017yhj,Ryu:2017qzn}. The increase slows down at $p_T >$ 2.5 GeV/$c$, the total $v_2(p_T)$ reaches a peak value at around $p_{T{\rm meson}}\sim 3$ GeV/$c$ and $p_{T{\rm baryon}}\sim 4.5$ GeV/$c$ before decreasing at large $p_T$, where the anisotropy caused by the geometric anisotropy of the QGP fireball and the length dependence of parton energy loss is significantly smaller than that from hydro and parton coalescence. The mass ordering is also reversed starting at $p_T\approx 2.5$ GeV/$c$. The change of $p_T$ dependence and mass ordering of $v_2(p_T)$ are caused mainly by the interplay among hadrons from hydro, coalescence and fragmentation. Since the transition from hydro to coalescence and fragmentation occurs at a fixed $p_{Ts}$ for the effective constituent quarks, both the value and the peak position of the final $v_2$ for $p$ is higher than $\pi$ and $K$ in the intermediate $3 < p_T < 8$ GeV/$c$ region, resembling the approximate NCQ scaling~ \cite{Fries:2003vb,Fries:2003kq,Greco:2003xt,Greco:2003mm,Greco:2003vf,Molnar:2003ff,Hwa:2004ng}. Though more hadrons are produced from fragmentation than from coalescence in this intermediate $p_T$ region (see~Fig.~\ref{fig:spectraall}),  $v_2$ of $p$ ($\pi, K$) from coalescence is, however, about a factor of 4 (2, 3) larger than that from fragmentation (see Fig.~\ref{fig:v2ptcontribution}). Coalescence, therefore, still contributes significantly to the final $v_2$ of all hadrons. Without coalescence (dashed lines in Fig.~\ref{fig:v2pid1020}),  CoLBT-hydro underestimates the $v_2$ by up to a factor of 2 in this $p_T$ region. The hadronic cascade further pushes hadrons, especially baryons, toward higher $p_T$, increasing both $R_{AA}$ and $v_2$ of identified particles in agreement with the experimental data at intermediate $p_T$.
CoLBT-hydro also predicts a slightly larger $v_2$ of $K$ than $\pi$ in this region because of the enhanced thermal strangeness in parton coalescence.

\begin{figure}
  \includegraphics[scale=0.43]{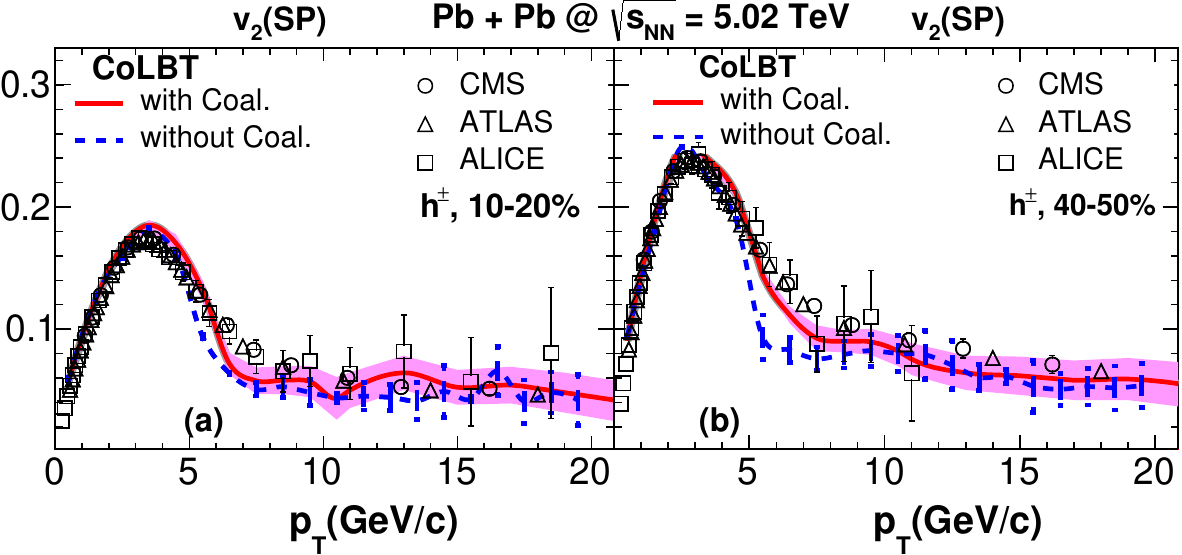}
  \caption{(Color online) CoLBT-hydro results on $v_{2}(p_T)$ of charged hadrons in (a) 10-20\% and (b) 40-50\% Pb+Pb collisions at $\sqrt{s_{NN}}=$5.02 TeV with (solid) and without quark coalescence (dashed) compared with experimental data~\cite{Sirunyan:2017pan,Aaboud:2018ves,Acharya:2018lmh}. }
  \label{fig:v2ptall}
\end{figure}

Finally, we show $v_2 (p_T )$ of all charged hadrons in 10-20\% and 40-50\% Pb+Pb collisions as compared to experimental data \cite{Sirunyan:2017pan,Aaboud:2018ves,Acharya:2018lmh} in Fig.~\ref{fig:v2ptall}.  At high $p_T>10$ GeV/$c$, fragmentation dominates the hadron production, $v_2$ for all light flavor hadrons becomes the same and arises from the geometric anisotropy of the medium and length dependence of the parton energy loss. 
Since the radiative parton energy loss $\Delta E$ has a logarithmic energy dependence and given the power-law behavior $1/p_T^{n}$ of the initial jet spectra, 
$R_{AA}\approx 1-n\Delta E/p_T$ 
eventually becomes 1 and $v_2$ approaches to 0 at high $p_T$. LBT without thermal-shower coalescence can describe this trend well \cite{Cao:2017umt,Cao:2017hhk}. CoLBT-hydro with the Hydro-Coal-Frag hadronization and hadron cascade can, therefore, consistently describe both $R_{AA}(p_T)$ and $v_2 (p_T )$ of charged hadrons in the whole $p_T$ range. As shown in Figs.~\ref{fig:raa} and \ref{fig:v2ptall}, CoLBT-hydro without quark coalescence significantly underestimates both $R_{AA}$ and $v_2(p_T)$ at the intermediate $p_T$ region.  We have carried out similar studies in other collisions systems at both RHIC and LHC energies. The combined approach in CoLBT-hydro model can also describe the colliding energy dependence well. These studies with careful analyses of model uncertainties will be presented in a future publication.

\noindent {\it 6. Summary}:
We carried out the first study of hadron spectra in A+A collisions that combines the state-of-the-art CoLBT-hydro, the Hydro-Coal-Frag hybrid hadronization and hadron cascade. We demonstrated that the interplay between hydro freeze-out at low-$p_T$, parton coalescence at intermediate $p_T$ and fragmentation at high $p_T$ can simultaneously explain the nuclear modification $R_{AA}$, elliptic anisotropy $v_2$ of charged and identified hadrons and their flavor dependence in the full range of $p_T$. The long-standing $R_{AA}\otimes v_2$ puzzle is solved for the first time with the inclusion of quark coalescence in this coupled approach, which significantly increases $v_2$ in the intermediate $p_T$ region. The predicted splitting between $v_2$ of $K$ and $\pi$ in the intermediate $p_T$ region can serve as a precision test for the coalescence mechanism. Parton coalescence has been shown to be essential to explain the NCQ scaling of $v_2$ at intermediate $p_T$ irrespective of the details of the coalescence models \cite{Fries:2003vb,Fries:2003kq,Greco:2003xt,Greco:2003mm,Greco:2003vf,Molnar:2003ff,Hwa:2004ng,Adams:2004bi,Adams:2005zg,Adare:2006ti,Abelev:2007ra,Adare:2012vq,Adare:2011vy}. This should also be the case for our combined approach with Hydro-Coal-Frag hybrid hadronization and hadron cascade to solve the $R_{AA}\otimes v_2$ puzzle.

\noindent {\it Acknowledgements}: 
We thank S. Cao, Y. He, C.-M. Ko, L. Pang, G. Qin, H. Song and Z. Yang for discussions. This work was supported in part by NSFC under grant Nos. 11935007, 11221504, 11861131009 and 11890714, the Fundamental Research Funds for Central Universities in China, US DOE under grant No. DE-AC02-05CH11231, US NSF under grant Nos. ACI-1550300 and OAC-2004571, EU ERDF and H2020 grant 82409, ERC grant ERC-2018-ADG-835105, Spanish AEI grant FPA2017-83814-P and MDM- 2016-0692, Xunta de Galicia Research Center accreditation 2019-2022 and the UCB-CCNU Collaboration Grant. Computations are performed at NSC3/CCNU.

\bibliography{bibliography}

\end{document}